# Recent reproducibility estimates indicate that negative evidence is observed over 30 times before publication

Michael Ingre
michael.ingre@su.se
Department of Psychology
Stockholm University

**Abstract**

The Open Science Collaboration recently reported that 36% of published findings from psychological studies were reproducible by independent researchers. We can use this information together with Bayes' theorem to estimate the statistical power needed to produce these findings under various assumptions and calculate the expected distribution of positive and negative evidence for a range of prior probabilities of the tested hypotheses; and by comparing this distribution to other findings indicating that >90% of publications in the psychological literature are statistically significant in support of the authors' hypothesis, we can estimate the magnitude of publication bias. The results indicate that negative evidence was observed 30--200 times before one was published.

*Introduction*

The Open Science Collaboration (OSC) recently reported that 36% of published positive (i.e. statistically significant) findings in three top psychological journals were reproducible by independent researchers (Open Science Collaboration, 2015). This finding is interesting in itself as an indicator of the veracity of published findings in psychology; but when used together with evidence from other studies indicating that the psychological literature is dominated by >90% significant findings supporting the authors hypothesis (Fanelli, 2010; T. D. Sterling, Rosenbaum, & Weinkam, 1995; Theodore D. Sterling, 1959) and Bayes' theorem, it can also help us to better understand statistical power and publication bias in psychological research.

*Introducing Bayes' theorem*

The analysis presented below makes use of Bayes' theorem and in one of its most basic and recognizable forms, it is used to calculate the conditional probability of an event A given event B:

$$P(A|B) = \frac{P(B|A)P(A)}{P(B)} \qquad \text{Eq (1)}$$

The theorem forms the fundamental basis for statistical inference since the two events (A & B) can be substituted with a specific scientific hypothesis (H) and the observed empirical evidence (E):

$$P(H|E) = \frac{P(E|H)P(H)}{P(E)} \qquad \text{Eq (2)}$$

The numerator of the equation describes the probability of observing the evidence when the hypothesis is true and the denominator describes the overall probability of observing the evidence (true or false). Thus, the theorem calculates the posterior probability of the hypothesis under investigation, after the evidence has been observed. The theorem can also be written using common statistical notation to determine the probability that a "statistically significant finding" describes a true finding rather than a type-1 error. The numerator in this case is simply the prior probability of the hypothesis $\theta$ multiplied by the statistical power $(1-\beta)$ of the test; the denominator includes the numerator and adds the type-1 errors ($\alpha$) when the hypothesis is false $(1-\theta)$:

$$\hat{\theta} = \frac{\theta(1-\beta)}{\theta(1-\beta) + \alpha(1-\theta)} \qquad \text{Eq (3)}$$

*Implications of >90% significant findings in the literature*

The psychological literature is dominated by >90% statistically significant findings supporting the author's hypothesis (Fanelli, 2010; T. D. Sterling et al., 1995; Theodore D. Sterling, 1959). Some simple math shows that these observations are only compatible with a very narrow set of prior probabilities and statistical powers if we want to assume an unbiased literature.

The prior ($\theta$) describes the probability that a hypothesis is true before data is observed and statistical power $(1-\beta)$ describes the probability that a study will confirm a true hypothesis; thus, the expected proportion of statistically significant findings supporting the researchers hypothesis is determined by the true findings $\theta(1-\beta)$, and the type-1 errors when the hypothesis is false $\alpha(1-\theta)$.

Even assuming that all tested hypotheses would be true *a priori* ($\theta = 1$) or that statistical power would be perfect ($1 - \beta = 1$) we find that we need a statistical power of at least $1 - \beta \geq .90$ or the prior to be $\theta \geq .90$ to observe >90% significant findings. Applying Bayes theorem (equation 3) and assuming a middle ground on both these parameters ($\theta = .95; 1 - \beta = .95$) we find that the posterior probability of such research is $\hat{\theta} \approx (.95 * .95)/(.95 * .95 + .05 - .05 * .95) \approx .997$; thus, approximately 99.7% of the published studies should describe true findings if we want to assume an unbiased literature.

*Expected reproducibility and a formal test of publication bias*

The next piece of the puzzle is the expected reproducibility rate from an unbiased literature with 99.7% true findings. Below we use $\hat{\theta}$ to denote the probability that a published finding is true calculated by Bayes' theorem in equation 3. The expected reproducibility ($r$) includes the true findings multiplied by the statistical power of the replication study $\hat{\theta}(1 - \beta)$ and the (presumably, small amount of) type-1 errors that is expected when the original finding would be false $\alpha(1 - \hat{\theta})$ given by equation 4:

$$r = \hat{\theta}(1 - \beta) + \alpha(1 - \hat{\theta})$$    Eq (4)

In replication studies using two-tailed test of a directed hypothesis, such as the one by the OSC (Open Science Collaboration, 2015), the nominal type-1 error rate is $\alpha = .025$. The expected reproducibility for unbiased research consistent with 90% significant findings is calculated using equation 4, assuming $1 - \beta = .95$ and $\hat{\theta} = .997$ as discussed above: $r = .997 * (1 - .05) + .025 * (1 - .997) = .947$; thus, approximately 94.7% of published findings would be reproduced by independent researchers.

A binomial test on the observed reproducibility of 36% (95% CI: 27%-46%; n=97) indicates strong evidence (p<.0000001) that the replication studies reported by the OSC were not drawn from an unbiased literature with 94.7% reproducibility; and thus, represent a formal test indicating publication bias in the psychological literature.

*Calculating statistical power and publication bias from the observed reproducibility*

The above calculations provide evidence that the literature is biased; but to determine the magnitude of bias we need to know the prior probability and statistical power of the original research. To assess these parameters we can form a system of equations where equation 3 is used to calculate the probability of the findings from the original studies to be true and equation 4 is used the calculate the expected reproducibility:

$$\hat{\theta} = \frac{\theta(1-\beta_3)}{\theta(1-\beta_3)+\alpha_3(1-\theta)}$$

$$r = \hat{\theta}(1-\beta_4) + \alpha_4(1-\hat{\theta}) \qquad \text{Eq (5)}$$

If we assume a nominal type-1 error rate of $\alpha_3 = .05$ for the original studies and a type-1 error rate of $\alpha_4 = .025$ for the replication studies as discussed above, an observed reproducibility of $r = .36$ as reported by OSC and statistical power to be identical in both original studies and the replication studies (i.e. $\beta_3 = \beta_4$), the system of equations given by equation 5 becomes solvable to find the statistical power ($1-\beta$) needed to produce the findings for a range of priors ($\theta$). Solving this system of equations can be complicated and I recommend using a computerized equation solver if there is a need to adapt the solution to different assumptions and then cross check the math to make sure the solution is correct (see the supplemental material for an example of how to do this).

Since the replication studies were larger than the original studies (Open Science Collaboration, 2015), and thus, is assumed to have larger statistical power than the original studies, we cannot assess statistical power of the original studies and replication studies directly; but we can define the two extremes of a range in which the true statistical power of the original studies must fall. The bottom end of this range assumes statistical power to be identical in the replication and original studies (i.e. $\beta = \beta_4 = \beta_3$) with the solution to $\beta$ given by equation 6 below; at the other extreme, statistical power is assumed to be perfect in the replication studies ($\beta_4 = 0$), and $\beta_3$ is given by equation 7 below:

$$\beta = \frac{39}{40} - \frac{67}{200\hat{\theta}} \qquad \text{Eq (6)}$$

where $\hat{\theta} = \frac{\sqrt{2\psi\theta} + 31\theta - 2\sqrt{2\psi}}{5(\theta-2)}$ and $\psi = \sqrt{\frac{\theta(313\theta + 335)}{(\theta-2)^2}}$

$$\beta_3 = \frac{2627\theta - 67}{2560\theta} \qquad \text{Eq (7)}$$

Now that we have estimates of $\beta_3$ for a range of priors, and thus, estimates of the statistical power of the original studies, we can calculate the expected probability of positive evidence $e_+ = \theta(1-\beta) + \alpha(1-\theta)$ and negative evidence $e_- = 1 - e_+$ for a range of plausible priors ($\theta$); and assuming a $.10/.90$ distribution of non-significant/significant findings in the literature as discussed above, we can also calculate publication bias $\varsigma = (e_-/e_+)/(.10/.90)$ indicating the number of times negative evidence was expected to be observed before such evidence was published in relation to the publication rate of positive evidence.

*Assessing plausible priors*

We cannot know the true priors in psychological research, and they will vary depending on the specific field; but for science and psychology in general we can probably exclude the highest and lowest priors as unrealistic. Ioannidis suggested priors in the range $.001 \leq \theta \leq .67$ for various fields and designs (Ioannidis, 2005) and his predictions are consistent with recent replication studies (Begley & Ellis, 2012; Ioannidis, Tarone, & McLaughlin, 2011; Open Science Collaboration, 2015; Prinz, Schlange, & Asadullah, 2011; Steward, Popovich, Dietrich, & Kleitman, 2012) suggesting that they were probably in right order of magnitude. In the present paper we restricted our analysis to priors in the range $.10 \leq \theta \leq .50$ since it is unlikely that the massively explorative research suggested by the smallest priors or the confirmatory research indicated by the largest priors are representative for the data we are analysing.

*Summary of findings*

Figure 1 presents the estimated range of publication bias ($\varsigma$) between the two extremes of replication power described by equations 6 & 7; and since the replication studies were larger than the original studies, we expect statistical power to also be larger, and the true value to fall between these two estimates. The findings indicate that negative evidence was expected to be observed in the range of 31--227 times before publication, assuming $\alpha = .05$. This reflects an expected proportion of positive findings of 4%--22%, a statistical power of 3%--60% in the original studies and a posterior probability of 34%--89% depending on the assumed prior (table 1). Assuming a higher reproducibility rate, at the upper end of the 95% confidence interval reported by the OSC of $r = .46$, suggest that negative evidence was observed 23-190 times before publication, indicating that these findings are robust to uncertainty in the reproducibility estimate due to chance (see table 2 in the supplemental material based on equation 10 & 11).

If we assume that other sources of bias were inflating the observed type-1 error rate of the original studies from its nominal value of $\alpha = .05$, it is possible to explain some of the excess significant findings observed in the literature. However, even assuming the nominal type-1 error rate to be inflated four times to $\alpha = .20$ in the original studies, we find that that negative evidence was still expected to be observed 17-50 times for each published finding (see table 1 using equations 8-9 in the supplementary material). Since we are analyzing *direct* replications, most methodological biases that may have been present in the original study, for example, from the design, procedure, measurements or the target group of subjects, should have been reproduced by the replication study; and thus, cannot explain the excess significant findings in the published literature.

*P-hacking*

A plausible candidate of bias that was not reproduced in the replication studies is p-hacking; Simmons et al (Simmons, Nelson, & Simonsohn, 2011) showed that p-hacking can bring the type-1 error rate up to 60% in a typical psychological study. However, p-hacking can also be seen as a special case of publication bias, since many negative findings are expected to be observed but never published; and we would not have made the same inferences if these negative observations had been reported. Thus, it is questionable if assuming an inflated type-1 error rate to adjust for p-hacking bias is a meaningful exercise, since the publication bias estimate ($\varsigma$) already accounts for suppressing non-significant findings.

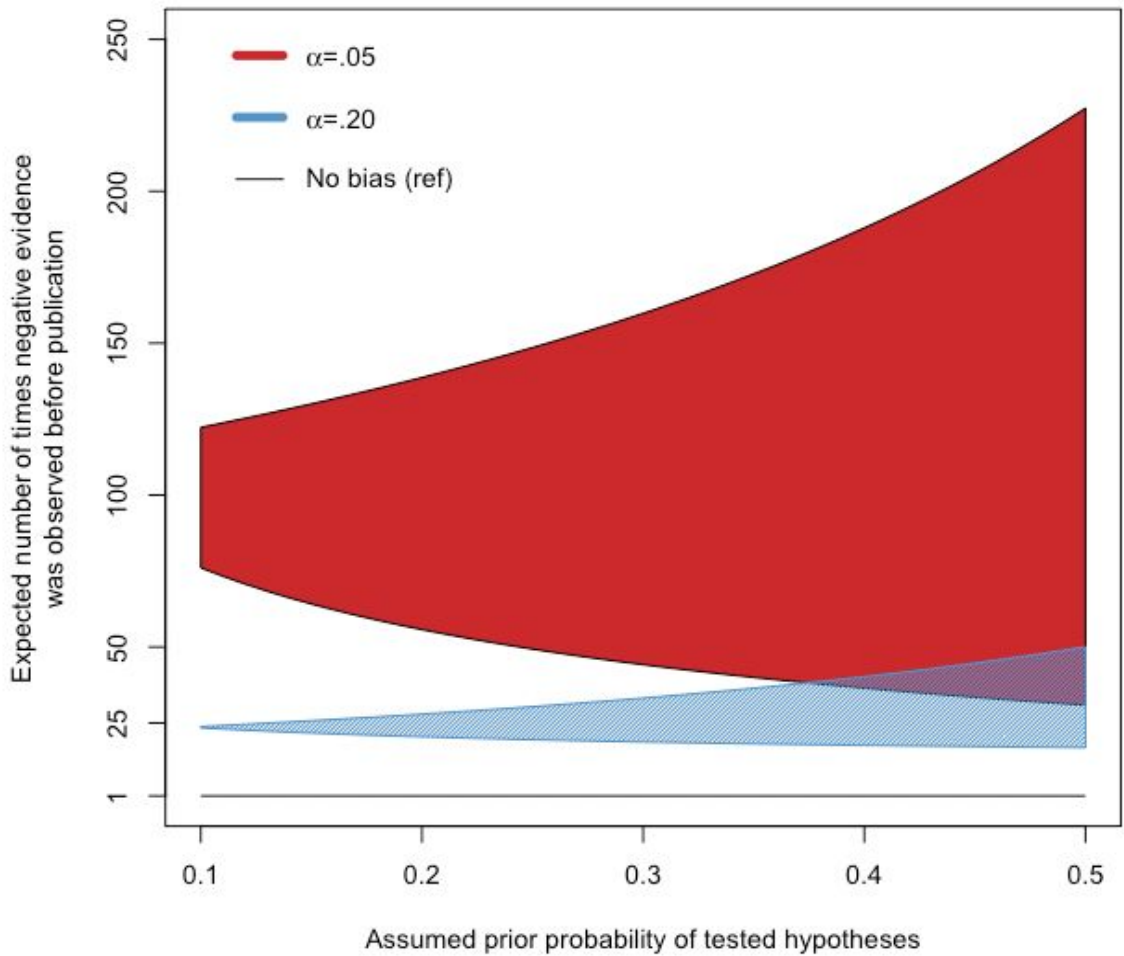

**Figure 1.** Estimated publication bias expressed as the expected number of times negative evidence was observed before publication, for prior probabilities in the range $.10 \leq \theta \leq .50$ under the assumption of an effective type-1 error rate of $\alpha = .05$ and $\alpha = .20$ in the original studies, and $\alpha = .025$ in the replication studies, a reproducibility rate of $r = .36$ and a literature with 90% positive findings.

Table 1: Publication bias in the psychological literature

| Assumed type-1 error rate of original studies | Assumed statistical power of replication studies | Assumed prior probability of tested hypotheses | Statistical power of original studies | Posterior probability of original studies | Expected proportion of positive findings | Publication bias |
|---|---|---|---|---|---|---|
| 0.05 | 0.608 | 0.1 | 0.608 | 0.575 | 0.106 | 76 |
| 0.05 | 0.495 | 0.2 | 0.495 | 0.712 | 0.139 | 56 |
| 0.05 | 0.447 | 0.3 | 0.447 | 0.793 | 0.169 | 44 |
| 0.05 | 0.420 | 0.4 | 0.420 | 0.848 | 0.198 | 36 |
| 0.05 | 0.402 | 0.5 | 0.402 | 0.889 | 0.226 | 31 |
|  |  |  |  |  |  |  |
| 0.05 | 1 | 0.1 | 0.236 | 0.344 | 0.069 | 122 |
| 0.05 | 1 | 0.2 | 0.105 | 0.344 | 0.061 | 139 |
| 0.05 | 1 | 0.3 | 0.061 | 0.344 | 0.053 | 160 |
| 0.05 | 1 | 0.4 | 0.039 | 0.344 | 0.046 | 188 |
| 0.05 | 1 | 0.5 | 0.026 | 0.344 | 0.038 | 227 |
|  |  |  |  |  |  |  |
| 0.2 | 0.977 | 0.1 | 0.977 | 0.352 | 0.278 | 23 |
| 0.2 | 0.728 | 0.2 | 0.728 | 0.476 | 0.306 | 20 |
| 0.2 | 0.614 | 0.3 | 0.614 | 0.568 | 0.324 | 19 |
| 0.2 | 0.545 | 0.4 | 0.545 | 0.645 | 0.338 | 18 |
| 0.2 | 0.495 | 0.5 | 0.495 | 0.712 | 0.348 | 17 |
|  |  |  |  |  |  |  |
| 0.2 | 1 | 0.1 | 0.942 | 0.344 | 0.274 | 24 |
| 0.2 | 1 | 0.2 | 0.419 | 0.344 | 0.244 | 28 |
| 0.2 | 1 | 0.3 | 0.244 | 0.344 | 0.213 | 33 |
| 0.2 | 1 | 0.4 | 0.157 | 0.344 | 0.183 | 40 |
| 0.2 | 1 | 0.5 | 0.105 | 0.344 | 0.152 | 50 |

Note: The table assumes a reproducibility rate of $r = .36$ using a two-tailed directed test with an effective type-1 error rate of $\alpha = .025$. Publication bias indicate the number of times negative evidence was expected to be observed before a finding was published assuming a literature with 90% positive findings.

*Discussion*

The present study used the recently reported reproducibility rate in psychology of 36% (Open Science Collaboration, 2015) together with information from a long series of reports indicating that >90% of published findings in psychology are statistically significant (Fanelli, 2010; T. D. Sterling et al., 1995; Theodore D. Sterling, 1959), to estimate publication bias based on the known mathematical relations described by Bayes' theorem. The findings suggest that negative evidence was observed in the range of 30-200 times before one was published, depending on the assumed prior probability of the research and statistical power of the replication studies.

Using Bayes' theorem in this way has some important implications: It assumes that hypotheses are either true or false; and such binary hypothesis testing has been criticized (Cohen, 1994). Indeed, it can be argued that there are no true non-zero associations associated with hypotheses on *observational* data. If we assume that no associations are truly zero, but are not interested in making inferences from very small true effect sizes, p-values from traditional (non-directional) null hypothesis significance testing (NHST) would be biased with inflated type-1 errors. In addition, we may conclude that any (non-directional) hypothesis is necessarily true, giving a trivial prior of $\theta = 1$. However, we should recognize that these are not limitations of binary hypothesis testing per se, but rather limitations of how specific hypotheses are formulated and tested. It is possible to define an alternative "NULL" hypothesis with a mean other than zero, to protect inferences from true effect sizes of "trivial" magnitudes (Ingre, 2013) and make the prior more informative in observational studies at $\theta < 1$. Also, binary NHST is not inherently problematic in true experimental designs since we can assume that the NULL is indeed zero. In the present study, we sought to estimate publication bias as a function of the statistical significance of tested hypotheses observed by the publishing authors, and the limitations discussed above applies similarly to how they would apply to the original studies.

The most crucial estimate used in the above analysis is the reported reproducibility rate of 36%. Reproducibility is a complicated concept with many different facets, in particular in psychology and the social sciences; some "true" findings may not be possible to replicate in a different time, social or cultural context; because the underlying meaning of the constructs used to design the study or define the variables may have changed. The underlying theory may still be valid but needs to be adapted to the new environment, and this has been proposed as an argument against the validity of direct replication of a study's methods on an independent sample (Stroebe

& Strack, 2014). But from a more general scientific perspective, it can be seen as a flaw in the formulated theory and the methods defined to test it; science needs to be verifiable to stand out from other types of claims and should have some generalizability to be a useful source of knowledge. If a presented theory or finding is very context-dependent and this information was lacking from the report: Can we still call it science? Another factor to consider is poorly described methods in the original study that may impact the success rate in replications; but this is essentially the same problem. If the study report did not present sufficient information to accurately replicate the methods: how can it be properly understood and evaluated by the readers?

Reproducibility may have been impaired because of mistakes made by the replicating team of researchers; though, this does not seem to be a large risk in the OSC study. The study was pre-registered and performed by well motivated researchers under more or less public scrutiny; the team were in frequent contact with authors of the original studies to obtain material and information about the design and procedure of their studies; and they employed a system of internal reviews of all studies to ensure quality. However, any potential mistakes that may have lowered the reproducibility rate is part of the overall type-2 error rate ($\beta$) in equation 4; and can be seen as a reduction of "statistical power" in the replication studies below what we have nominally assumed.

An important assumption was that statistical power was higher in the replication studies than in the original studies. This allowed us to define the extremes of a range in which the true value must fall. This assumption seems to hold at the aggregate level, with a reported median df=68 for the replication studies and df=54 for the original studies (Open Science Collaboration, 2015). However, the sample size difference between original studies and replication studies was not that large, leaving a small margin for potential mistakes that may have reduced power in the attempted replications. In addition, the OSC choose to design their replications with a target statistical power of 90% based on the originally reported effect size, instead of explicitly designing each replication with higher power than the original study. The data downloaded from their github repository shows that only 70% of the replications were designed with a larger sample than the original, 10% had the same sample size, and 20% were actually smaller than the original study. Thus, while there is no information to suggest that our assumption does not hold at the aggregated level, statistical power cannot be assumed to be substantially larger in the replication studies, and the true values of all estimates reported in the present study are likely to fall closer to the reported lower end of statistical power than at the higher end; and there is some added uncertainty due to heterogeneity in statistical power of the replication studies versus the original studies.

Studies eligible for replication by the OSC were picked from three top journals in psychology and approximately one third of the total sample was never submitted for replication, mostly because these studies were deemed infeasible to replicate e.g. because they required special samples, knowledge or equipment. This introduces uncertainty and potential bias in the reproducibility estimate; it is not inconceivable that the more specialized or complicated designs would have worse (but less likely better) reproducibility. Thus, the reproducibility rate estimated by the OSC is an estimate representative of the two thirds most accessible research in three high quality psychological journals; and might not generalize to psychology in general.

For the studies replicated by OSC we have tentative estimates of previously unattainable properties of psychological science, suggesting a posterior probability in the range $.34 < \hat{\theta} < .89$ as the result of statistical power of the original studies of $.03 < 1 - \hat{\beta} < .60$ depending on the assumed prior and replication statistical power. We should keep in mind that these are estimates of a range with implausible extremes and only indicates the area, together with the assumed prior as illustrated in figure 1, in which the true value must fall when our assumptions hold. These findings also needs to be interpreted keeping in mind the limitations discussed above for binary hypothesis testing and need to consider any potential methodological biases in the original studies that may have been replicated by the OSC; and thus, describe true associations in data but not necessary provide evidence for the suggested theory or hypothesis. However, they do suggest that published findings may have a better veracity than what was implied by the 36% reproducibility reported by OSC due to much poorer statistical power in the replication studies than what was assumed by the OSC.

While the findings presented in this paper refer to psychological studies, data available from other scientific fields indicates slightly less pronounced focus on positive evidence with 70-90% significant findings supporting the authors hypothesis (Fanelli, 2010; T. D. Sterling et al., 1995) but even worse reproducibility rates in the range 11-24% in certain fields (Begley & Ellis, 2012; Prinz et al., 2011; Steward et al., 2012), suggesting that the findings presented here might be generalizable to a broader area of science. However, specific fields with a higher proportion of published negative evidence and/or a higher demonstrated reproducibility are likely to be less affected by publication bias.

One should recognize that most suppressed findings describe NULL effects that many may find uninformative or not interesting (Stroebe & Strack, 2014); but the fact that they are never published makes it more likely that similar studies are performed repeatedly by independent researchers; and eventually one will become significant by chance, dramatically increasing its chance of being published. Thus, the fact that such a large portion of negative evidence was suppressed from publication not only represents a serious threat to the veracity of published *positive* evidence; it also means that researchers are likely to spend time and resources testing hypotheses that should already have been rejected.

Publication bias may be the single most important problem to solve in science in order to increase the efficiency of the scientific project and bring the veracity of published research to higher standards. The implications of suppressing >30 negative findings for each one published should not be underestimated since with $\alpha = .05$, we expect a significant finding by chance for every 20 observations made on random data: even when studied associations are truly NULL we expect the literature to be dominated by significant findings.

# References

Begley, C. G., & Ellis, L. M. (2012). Drug development: Raise standards for preclinical cancer research. *Nature*, *483*(7391), 531–533.

Cohen, J. (1994). The earth is round (p < .05). *The American Psychologist*, *49*(12), 997.

Fanelli, D. (2010). "Positive" results increase down the Hierarchy of the Sciences. *PloS One*, *5*(4), e10068.

Ingre, M. (2013). Why small low-powered studies are worse than large high-powered studies and how to protect against "trivial" findings in research: comment on Friston (2012). *NeuroImage*, *81*, 496–498.

Ioannidis, J. P. A. (2005). Why most published research findings are false. *PLoS Medicine*, *2*(8), e124.

Ioannidis, J. P. A., Tarone, R., & McLaughlin, J. K. (2011). The false-positive to false-negative ratio in epidemiologic studies. *Epidemiology* , *22*(4), 450–456.

Open Science Collaboration. (2015). Estimating the reproducibility of psychological science. *Science*, *349*(6251). http://doi.org/10.1126/science.aac4716

Prinz, F., Schlange, T., & Asadullah, K. (2011). Believe it or not: how much can we rely on published data on potential drug targets? *Nature Reviews. Drug Discovery*, *10*(9), 712.

Simmons, J. P., Nelson, L. D., & Simonsohn, U. (2011). False-positive psychology: undisclosed flexibility in data collection and analysis allows presenting anything as significant. *Psychological Science*, *22*(11), 1359–1366.

Sterling, T. D. (1959). Publication Decisions and Their Possible Effects on Inferences Drawn from Tests of Significance--Or Vice Versa. *Journal of the American Statistical Association*, *54*(285), 30–34.

Sterling, T. D., Rosenbaum, W. L., & Weinkam, J. J. (1995). Publication Decisions Revisited: The Effect of the Outcome of Statistical Tests on the Decision to Publish and Vice

Versa. *The American Statistician*, *49*(1), 108–112.

Steward, O., Popovich, P. G., Dietrich, W. D., & Kleitman, N. (2012). Replication and reproducibility in spinal cord injury research. *Experimental Neurology*, *233*(2), 597–605.

Stroebe, W., & Strack, F. (2014). The Alleged Crisis and the Illusion of Exact Replication. *Perspectives on Psychological Science: A Journal of the Association for Psychological Science*, *9*(1), 59–71.

# Supplementary material

R-code to replicate all findings can be found here:
https://github.com/micing/reproducibility_bias

*Solving the system of equations to find $\beta$*

The system of equations defined by equation 5 in the main text and replicated below needs to be solved for unique values of the reproducibility ($r$) and the type-1 error rate of the original ($\alpha_3$) and replication studies ($\alpha_4$). This can be simplified using a computerized equation solver and cross checking the math for the suggested solution. Code to cross check the math is available at the github repository listed above and below are links to one web based equation solver giving the solution presented by equation 6 & 7 in the main text ($r = .36, \alpha_3 = .05, \alpha_4 = .025$):

$$\hat{\theta} = \frac{\theta(1-\beta_3)}{\theta(1-\beta_3) + \alpha_3(1-\theta)}$$
$$r = \hat{\theta}(1-\beta_4) + \alpha_4(1-\hat{\theta}) \qquad \text{Eq (5)}$$

Equation 6 (assuming $\beta = \beta_3 = \beta_4$):
https://www.wolframalpha.com/input/?i=r%3D.36,+r%3Dp*(1-beta)%2B(1-p)*.025,+p%3D+(theta*(1-beta))%2F+(theta*(1-beta)+%2B+alpha*(1-theta)),+1%3Etheta%3E0,+1%3Ep%3E0,+alpha%3D.05,++solve+beta&f=1

Solution for $\hat{\theta}$
https://www.wolframalpha.com/input/?i=theta%3D(10*p%5E2)%2F(5*p%5E2-62*p%2B67),+0%3Ctheta%3C1,+0%3Cp%3C1,+solve+p&f=1

Equation 7 (assuming $\beta_4 = 0$):
https://www.wolframalpha.com/input/?i=r%3D.36,+r%3Dp*(1-0)%2B(1-p)*.025,+p%3D+(theta*(1-beta))%2F+(theta*(1-beta)+%2B+alpha*(1-theta)),+1%3Etheta%3E0,+1%3Ep%3E0,+alpha%3D.05,++solve+beta&f=1

Note: $\hat{\theta}$ has been substituted with "p" because of limitations in the software and make sure to request the "exact form" to get a solution identical to the one presented in the paper.

***Equations for $\beta$ assuming $r = .36$, $\alpha_3 = .20$ and $\alpha_4 = .025$.***

These equations are similar to equation 6 & 7 but assume $\alpha_3 = .20$. Equation 5 assumes power to be identical in the original studies and replication studies. Equation 6 assumes 100% power in the replication study.

$$\beta = \frac{39}{40} - \frac{67}{200\hat{\theta}} \qquad \text{Eq (8)}$$

where $\hat{\theta} = \dfrac{14\sqrt{2}\psi\theta + 31\theta - 16\sqrt{2}\psi}{5(7\theta - 8)}$ and $\psi = \sqrt{-\dfrac{\theta(173\theta - 335)}{(7\theta - 8)^2}}$

$$\beta_3 = \frac{707\theta - 67}{640\theta} \qquad \text{Eq (9)}$$

***Equations for $\beta$ assuming $r = .46$, $\alpha_3 = .20$ and $\alpha_4 = .025$.***

These equations are similar to equation 6 & 7 but assume $r = .46$ matching the upper end of the 95% confidence interval reported by the Open Science Collaboration. Equation 7 assumes power to be identical in the original studies and replication studies. Equation 8 assumes 100% power in the replication study.

$$\beta = \frac{3(65\hat{\theta} - 29)}{200\hat{\theta}} \qquad \text{Eq (10)}$$

where $\hat{\theta} = \dfrac{\sqrt{2}\psi\theta + 41\theta - 2\sqrt{2}\psi}{5(\theta - 2)}$ and $\psi = \sqrt{\dfrac{\theta(623\theta + 435)}{5(\theta - 2)^2}}$

$$\beta_3 = \frac{749\theta - 29}{720\theta} \qquad \text{Eq (11)}$$

**Table 2:** Publication bias in the psychological literature assuming a reproducibility of r=.46

| Assumed type-1 error rate of original studies | Statistical power of replication studies | Prior probability of tested hypotheses | Statistical power of original studies | Posterior probability of original studies | Expected proportion of positive findings | Publication bias |
|---|---|---|---|---|---|---|
| 0.05 | 0.729 | 0.1 | 0.729 | 0.618 | 0.118 | 67 |
| 0.05 | 0.604 | 0.2 | 0.604 | 0.751 | 0.161 | 47 |
| 0.05 | 0.552 | 0.3 | 0.552 | 0.826 | 0.201 | 36 |
| 0.05 | 0.522 | 0.4 | 0.522 | 0.874 | 0.239 | 29 |
| 0.05 | 0.503 | 0.5 | 0.503 | 0.910 | 0.277 | 24 |
| | | | | | | |
| 0.05 | 1 | 0.1 | 0.363 | 0.446 | 0.081 | 102 |
| 0.05 | 1 | 0.2 | 0.161 | 0.446 | 0.072 | 116 |
| 0.05 | 1 | 0.3 | 0.094 | 0.446 | 0.063 | 133 |
| 0.05 | 1 | 0.4 | 0.060 | 0.446 | 0.054 | 157 |
| 0.05 | 1 | 0.5 | 0.040 | 0.446 | 0.045 | 190 |

**Note:** The table assumes a reproducibility rate of r=.46 and a type-1 error rate of alpha=.025 (two tailed test) in the replication studies. Publication bias indicate the number of times negative evidence is observed before a finding is published assuming a literature with 90% positive findings.